\documentstyle[aps,prl,floats,epsfig]{revtex}

\begin{document}

\title{Spectrum of neutrino masses and their nature in the light of
present and future experiments.}
\author{M. Czakon$^a$, J. Studnik$^a$ and M. Zra{\l}ek$^{a,b}$}
\address{$^a$ Department of Field Theory and Particle Physics, Institute 
of Physics, University of Silesia, Uniwersytecka 4, PL-40-007 Katowice, Poland \\
$^b$ Depto. de Fisica Teorica y del Cosmos, Universidad de Granada,
E-18071 Granada, Spain}
\maketitle

\begin{abstract}
The present experimental data on neutrino oscillations, neutrinoless double beta decay and tritium beta decay
are collected together and possible mass ranges for Dirac and Majorana neutrinos are found. Four future 
experimental situations are investigated: both decay experiments give only upper bounds, one of them
gives a positive result ($|\langle m_\nu \rangle | \neq 0$ or $m_\beta \neq 0$), or finally both effective
neutrino masses are different from zero ($|\langle m_\nu \rangle | \neq 0$ and $m_\beta \neq 0$). Each
scenario gives new information on neutrino masses and nature but only the last has a chance to resolve
the problem and give some additional information on $CP$ violation in the lepton sector.
\end{abstract}

\vspace{.5cm}

The problem of neutrino mass spectrum and its nature is the most important issue in the lepton part of
the Standard Model. What new information can we obtain from the last experimental results, and what are the
future perspectives? Three kinds of experiments play a fundamental role in answering this question. Two are
traditional and known for years: beta decay and neutrinoless double beta decay ($(\beta\beta)_{0\nu}$) of
nuclei. Already Fermi \cite{fermi} in 1934 and Furry \cite{furry} in 1939 realized
that both processes are important to find the
neutrino mass and nature. The third type constitute the neutrino
oscillation experiments \cite{review}. We strongly believe
that neutrino oscillations are responsible for anomalies observed in
solar \cite{davis}, atmospheric \cite{fukuda} and LSND
\cite{athanassopoulos} experiments.
There are trials of alternative explanations of the observations
\cite{pakvasa} but they require much more sophisticated
assumptions (as for example the breaking of the equivalence principle, breaking of 
the special theory of relativity, the
neutrino decay with life-time much below expectations or huge neutrino magnetic moments) and give much 
poorer fits to the data \cite{lipari}.

The end of the Curie plot in tritium beta decay has been observed since the late forties giving now
a bound on the effective electron neutrino mass of $m_\beta \leq 2.8
eV$ \cite{weinheimer} and $m_\beta \leq 2.5 eV$ \cite{lobashev} both
with $95\%$ c.l.. Although less pronounced, the problem of negative $m_\beta^2$ remains.

Trials of finding the neutrinoless double beta decay of even-even
nuclei have also been conducted for years. The
best result for the effective electron neutrino mass $|\langle m_\nu \rangle |$ bound comes from the last 
experiment with $^{76}Ge$. The Heidelberg-Moscow collaboration found that the half life time of $^{76}Ge$ is
bounded as
\begin{equation}
T^{0\nu}_{1/2} (Ge) > 5.7\times 10^{25} years,
\end{equation}
which gives a bound of $|\langle m_\nu \rangle | < 0.2 eV$ \cite{baudis}. The derivation of this bound required the
calculation of complicated nuclear matrix elements. A discrepancy of the order of a factor of 3 \cite{pdg} between
the independent studies has been found. As a consequence the uncertainty of $|\langle m_\nu \rangle |$
is of the order of $\sqrt{3}$. 

\begin{figure}[t]
\begin{center}
\epsfig{figure=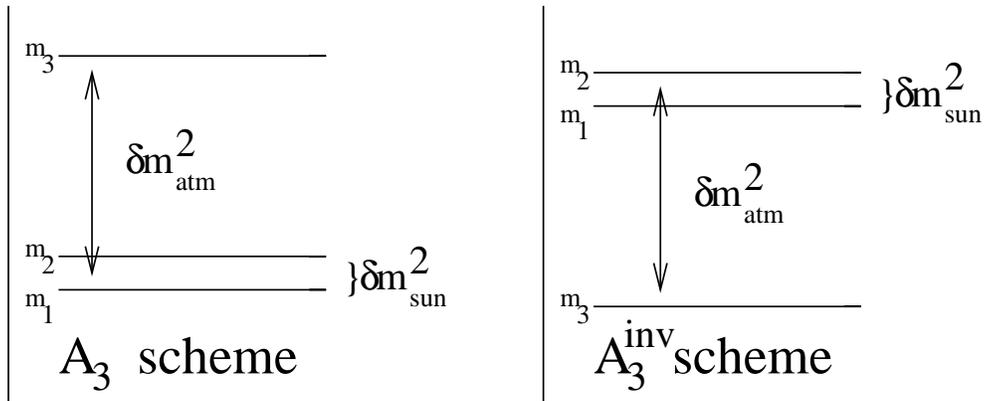, scale=1.4}
\caption{Two possible mass spectra which can describe the oscillation data. Scheme $A_3$, normal mass hierarchy,
has a small gap between $m_1$ and $m_2$ to explain the oscillation of solar neutrinos and a larger gap
for the atmospheric neutrinos. In the inverse mass hierarchy scheme $A_3^{inv}$, $m_3$ is the lightest neutrino
mass and $\delta m^2_{atm} = - \delta m^2_{31}$. Both schemes are not distinguishable by present experimental data
They will be discriminated by future experiments. \label{spectra}}
\end{center}
\end{figure}

Finally, the oscillation experiments give results on the basis of three
different observations: solar \cite{cleveland}, atmospheric \cite{fukuda2} and LSND \cite{athanassopoulos2}. 
The results of the last of these, if correct and
explained through neutrino oscillations requires the introduction of a fourth sterile light neutrino species.
The observation of the LSND collaboration has not been confirmed by KARMEN \cite{karmen} and is partially excluded by Bugey \cite{bugey}
and BNL776 \cite{bnl776} experiments. In such circumstances we will assume that only three massive neutrinos exist which
explain the solar and atmospheric neutrino anomalies. The case of four
massive neutrinos can be studied in a similar
way. Even with this assumption the oscillatory data has some ambiguities related in particular to the existence
of several possible solutions of the solar neutrino problem. The values of $\delta m^2$ and $\sin^2{2 \theta}$
for the atmospheric and the four solar solutions (SMA MSW, LMA MSW, LOW MSW and VO) of the observed anomalies 
are collected in Table~\ref{first}. As there are definitely two scales of $\delta m^2$, $\delta m^2_{atm} \gg
\delta m^2_{sol}$, two possible neutrino mass spectra must be considered (Fig.~\ref{spectra}). The first, known
as normal mass hierarchy ($A_3$) where $\delta m^2_{sol} = \delta m^2_{21} \ll \delta m^2_{32} \approx \delta
m^2_{atm}$ and the second, inverse mass hierarchy spectrum ($A^{inv}_3$) with $\delta m^2_{sol} = \delta 
m^2_{21} \ll \delta m^2_{atm} \approx -\delta m^2_{31}$. Both schemes are not distinguishable by present
experiments. There is hope that next long base line experiments (e.g. MINOS, ICANOE) will do that.
\begin{table}
\begin{tabular}{|c|cc|cc|}
Experiment & Allowed & range & Best & fits \\ 
solutions & $\delta m^2 [eV^2]$ & $\sin^2 2\theta$ & $\delta m^2 [eV^2]$ & $\sin^2 2\theta$ \\
\hline
Atmospheric neutrinos \cite{fukuda2,atmospheric}& $(1.5 - 7)\times 10^{-3}$ & $0.84-1$ & $3.5\times10^{-3}$ & $1.0$ \\
Solar neutrinos &&&& \\
MSW SMA \cite{hata,fogli}& $(4-10)\times 10^{-6}$ & $0.001-0.01$ & $5.2\times 10^{-6}$ & $0.0065$ \\
MSW LMA \cite{hata,fogli}& $(1.5-10)\times 10^{-5}$ & $0.59-0.98$ & $2.94\times 10^{-5}$ & $0.77$ \\
MSW LOW \cite{hata,fogli}& $(7-20)\times 10^{-8}$ & $0.68-0.98$ & $1.24\times 10^{-7}$ & $0.9$ \\
VO \cite{barger}& $(0.5-8)\times 10^{-10}$ & & $4.42 \times 10^{-10}$ & $0.93$ \\
\end{tabular}
\caption{The allowed ranges and best fit values of $\sin^2 2\theta$ and $\delta m^2$ for the atmospheric and different types
of solar neutrino oscillations \label{first}}
\end{table}

The values of the allowed range of $\delta m^2$ and $sin^2 {2\theta}$ in the Table~\ref{first} are presented
with a $95\% c.l.$. As we will see the value of $\sin^2{2\theta}$  plays a decisive role in our considerations.
A value of $\sin^2{2\theta}<1$ is crucial. If we look at the data \cite{hata}, we see that a maximal mixing is still 
possible at $99\% c.l.$ for three solutions of the solar neutrino problem LMA, LOW and VO \cite{fogli}. However, the
best fit values, contrary to the atmospheric neutrino case, are smaller than 1. Still, it is possible that
this is only a fluctuation and future better data are needed to solve this problem. Here we assume that the 
tendency observed in experimental data is real and future experiments
will confirm that $\sin^2{2\theta} <1$.

Two elements of the first row of the mixing matrix $|U_{e1}|$ and
$|U_{e2}|$ can be expressed by the third element $|U_{e3}|$ and the
$\sin^2{2\theta}$
\begin{equation}
|U_{e1}|^2 = (1-|U_{e3}|^2) \frac{1}{2} (1+\sqrt{1-\sin^2{2\theta}}),
\end{equation}
and
\begin{equation}
|U_{e2}|^2 = (1-|U_{e3}|^2) \frac{1}{2} (1-\sqrt{1-\sin^2{2\theta}}).
\end{equation}
The value of the third element $|U_{e3}|$ is not fixed yet and only
different bounds exist for it \cite{fogli2}. We will take the bound directly
inferred from the CHOOZ experiment \cite{apollonio}
\begin{equation}
|U_{e3}|^2 < 0.04.
\end{equation}
In spite of various complications and uncertainties even now we will
get some information about the neutrino mass spectrum and their
nature. We hope however, that such considerations with better data
will give a key to the solution of the problem in the future.

For the three massive neutrino scenario (without LSND) the oscillation
experiments give the lower bound on the highest mass of neutrinos
\begin{equation}
\label{bound}
(m_\nu)_{max} > \sqrt{\delta m^2_{atm}} \approx 0.06 eV
\end{equation}
and a bound on the absolute value of the difference of two masses
\begin{equation}
\label{difference}
|m_i - m_j| < \sqrt{\delta m^2_{atm}}.
\end{equation}
No upper bound on neutrino masses can be inferred from oscillation
experiments alone. So, in the schemes in Fig~\ref{spectra}, we only
know that $m_{3,2} \geq 0.06eV$, for $A_3$ and $A^{inv}_3$ schemes
respectively.

The tritium beta decay measures the effective electron neutrino mass
$m_\beta$
\begin{equation}
m_\beta = \left[ \sum_{i=1}^{3} |U_{ei}|^2 m_i^2 \right]^{1/2},
\end{equation}
and no upper bound on neutrino masses can be given.
We can only find that
\begin{equation}
(m_\nu)_{min} \leq m_\beta \leq (m_{\nu})_{max}
\end{equation}
where $(m_{\nu})_{min}$ denotes the lowest neutrino mass. In practice
this means that $(m_\nu)_{min} < 2.5 eV$, but $(m_\nu)_{max}$ can be
very large. Connecting both the tritium beta decay data and the
oscillatory data we can find the following upper bound on the neutrino
mass \cite{barger2}
\begin{equation}
\label{highest}
\sqrt{\delta m^2_{atm}} \leq (m_\nu)_{max} \leq
\sqrt{(m_\beta)^2+\delta m^2_{atm}}.
\end{equation}
This means that the mass of the heaviest neutrino must be somewhere in
between $0.06eV$ and $2.5eV$.

The estimations which we get up to now do not depend on the neutrino
nature. It is well known that the electron energy distribution in
nuclei beta decay and flavor oscillations do not distinguish Dirac
from Majorana neutrinos \cite{review2}. Therefore, the bounds (Eqs.~\ref{bound}, \ref{difference}
and Eq.~\ref{highest}) are valid for both neutrino types. This is not
the case of the neutrinoless double beta decay of nuclei. This decay
is only possible for a massive Majorana neutrino \cite{review3}. Neglecting all other
mechanisms which can participate in the process we can derive a bound
on
\begin{equation}
|\langle m_\nu \rangle | \equiv \left| \sum^{3}_{i=1} U^2_{ei} m_i \right|.
\end{equation}
Alone, this bound is of little value since it only means that
\begin{equation}
0\leq |\langle m_\nu \rangle | \leq (m_\nu)_{max}.
\end{equation}
Let us now consider however all three experiments together \cite{all}. If
neutrinos are Dirac particles then $|\langle m_\nu \rangle | = 0$ and
we do not have any additional information from $(\beta\beta)_{0\nu}$.
All we can say about masses of Dirac neutrinos follows from $^3_1 H$
decay
and oscillation experiments and is given by Eqs.~\ref{bound}, \ref{difference}
and Eq.~\ref{highest}. 

\begin{figure}[t]
\epsfig{figure=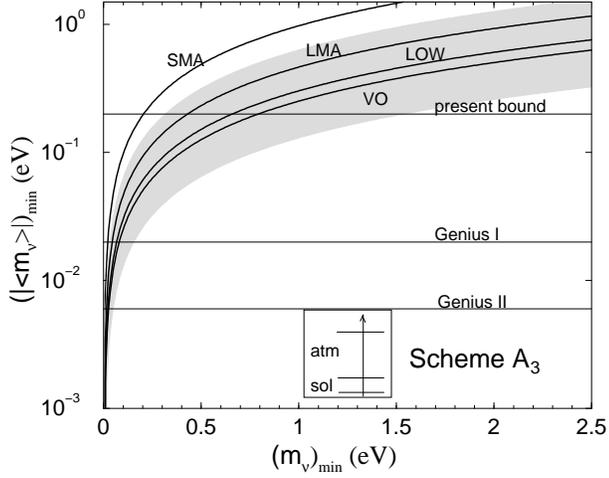,width=8cm} \\
\caption{Scheme $A_3$. The curves of $|\langle m_\nu \rangle |_{min}$
for the different solar neutrino oscillation solutions were obtained with
the values of $\sin^2 2\theta_{sun}$ quoted in the text and $|U_{e3}|^2=0.01$ . The shaded region is given by the largest range in 
Table~\ref{first}, that is the range for the LMA solution. For the SMA solution the $95\%c.l.$ range (.001-.01) is described 
by one curve in the present scale.
\label{a3}}
\end{figure}

\begin{figure}[t]
\epsfig{figure=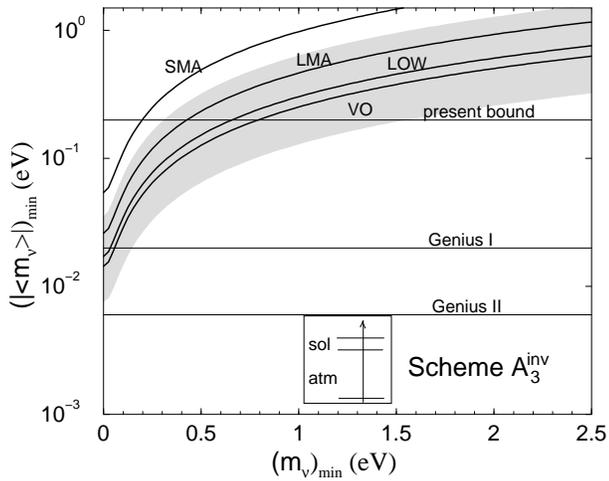,width=8cm} \\
\caption{Scheme $A_3^{inv}$. The curves were obtained with the same assumptions as in Fig.~\ref{a3}.
\label{a3inv}}
\end{figure}

If neutrinos are Majorana particles then the bound on $|\langle m_\nu
\rangle |$ works and we can find new restrictions. For three neutrinos
in the scheme $A_3$ (Fig.~\ref{spectra}) we have
\begin{eqnarray}
\label{basic}
|\langle m_\nu \rangle | &=& \left| |U_{e1}|^2 (m_\nu)_{min}+|U_{e2}|^2
e^{2i\phi_2} \sqrt{(m_\nu)_{min}^2+\delta m^2_{sol}} \right. \nonumber \\
&+& \left. |U_{e3}|^2 e^{2i\phi_3} \sqrt{(m_\nu)_{min}^2+\delta m^2_{sol}+\delta m^2_{atm}} \right| ,
\end{eqnarray}
and similarly in the case of the scheme $A^{inv}_3$. Three
parameters defined above are unknown, $(m_\nu)_{min}$ and the Majorana
$CP$ violating phases $\phi_2$ and $\phi_3$. We are not able to
predict the value of $|\langle m_\nu \rangle |$ but lower $|\langle
m_\nu \rangle |_{min}$
and upper $|\langle m_\nu \rangle |_{max}$ limits as function of
$(m_\nu)_{min}$ can be obtained. In Fig.~\ref{a3} (scheme $A_3$) and 
Fig.~\ref{a3inv} (scheme $A^{inv}_3$) $|\langle
m_\nu \rangle |_{min}$ is shown as function of $(m_\nu)_{min}$ for the
four possible solutions of the solar neutrino anomaly \cite{nasze1}. The shaded
regions give the uncertainties of the results caused by the allowed
region of input parameters (mostly $\sin^2{2\theta}$). For the SMA MSW
solution the $95\% c.l.$ range of $|\langle
m_\nu \rangle |_{min}$ is described by one curve in the adopted
logarithmic scale.
\twocolumn
\begin{figure}[t]
\epsfig{figure=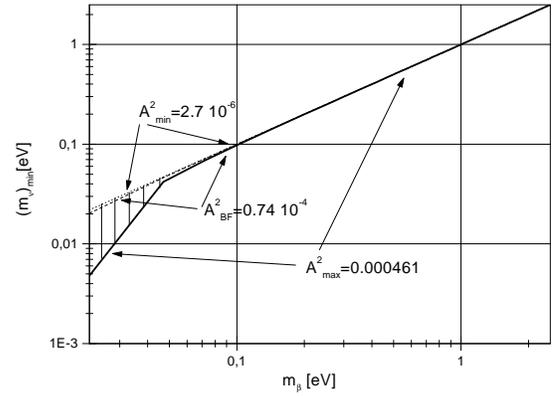,width=9cm}
\caption{$(m_\nu)_{min}$ as function of $m_\beta \approx \kappa'$
($(m_{\nu})_{min}=[\kappa'^{2}-(1-|U_{e1}|^{2})\delta m^{2}_{solar}-|U_{e3}|^{2}\delta m^{2}_{atm}]^{1/2}$, see Eq.~\ref{mass1}), 
with oscillation parameters changing in the range
specified in Table~\ref{first}. The solar neutrino solution is the LMA MSW. The difference is visible only for
$\kappa' < 0.1eV$\label{mnumin}}
\end{figure}
\onecolumn
{\bf Present Experimental Data}
\\ \\ 
We see that already with the present experimental
data the possible values of $|\langle
m_\nu \rangle |_{min}$ depicted in Fig.~\ref{a3} and Fig.~\ref{a3inv}
exceed the bound on $|\langle m_\nu \rangle |$ 
\begin{equation}
\label{test}
|\langle m_\nu \rangle |_{min}  > 0.2eV,
\end{equation}
for some values of $(m_\nu)_{min}$. This means that Majorana
neutrinos with masses above some $(m_\nu)_{min}$ are forbidden. The limiting
mass depends on the solution to the solar neutrino problem and on the
adopted value of $\sin^2{2\theta}$ and $|U_{e3}|^2$. For the SMA MSW solution in both
$A_3$ and $A^{inv}_3$ schemes $(m_\nu)_{min} = 0.22eV$. For LMA, LOW
and VO with $(\sin^2{2\theta})_{max} = 0.98$ and $|U_{e3}|^2=0.01$, $(m_\nu)_{min} = 1.5
eV$ \cite{nasze1,nasze2}.

\begin{figure}[t]
\epsfig{figure=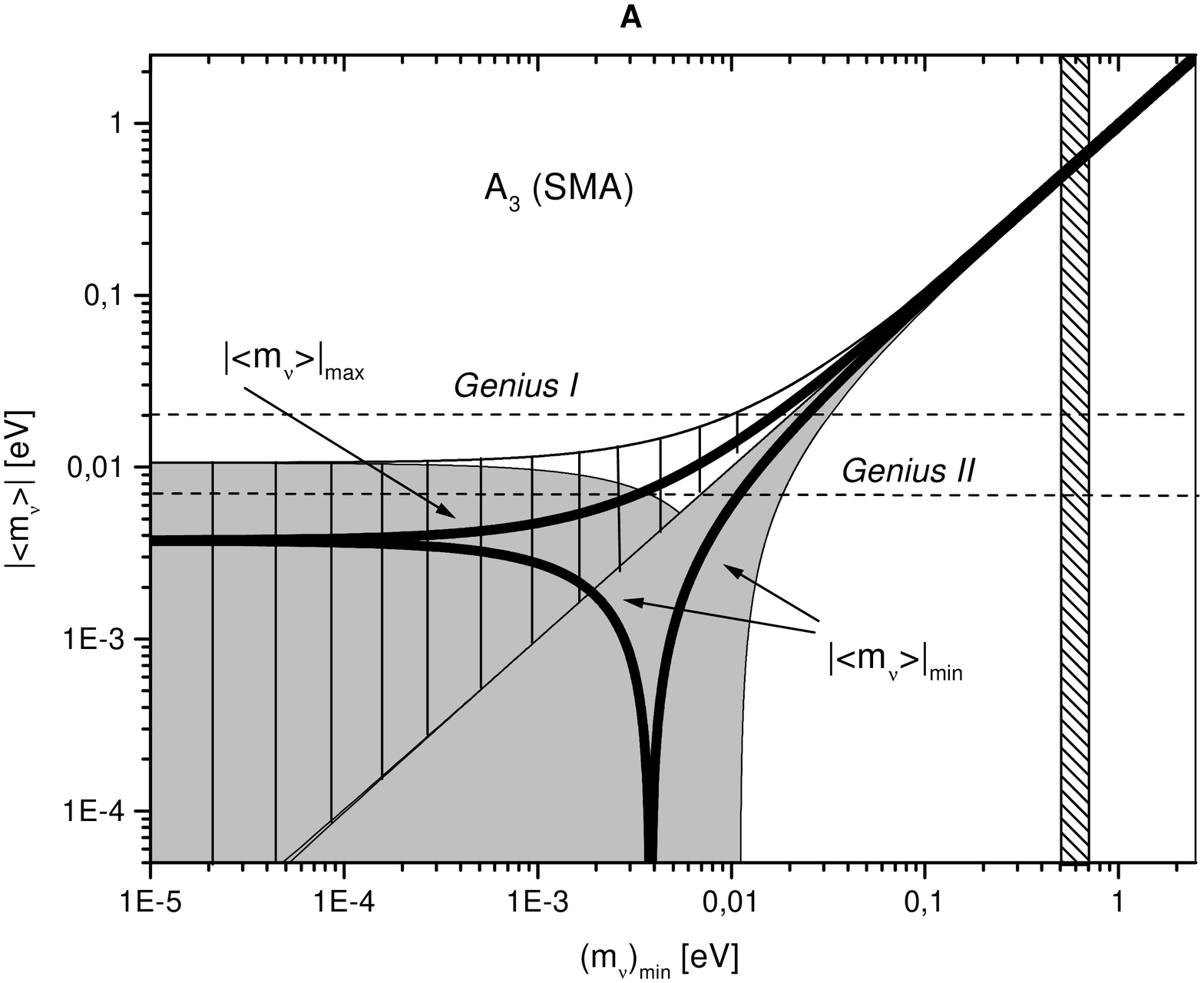,width=8.5cm}
\epsfig{figure=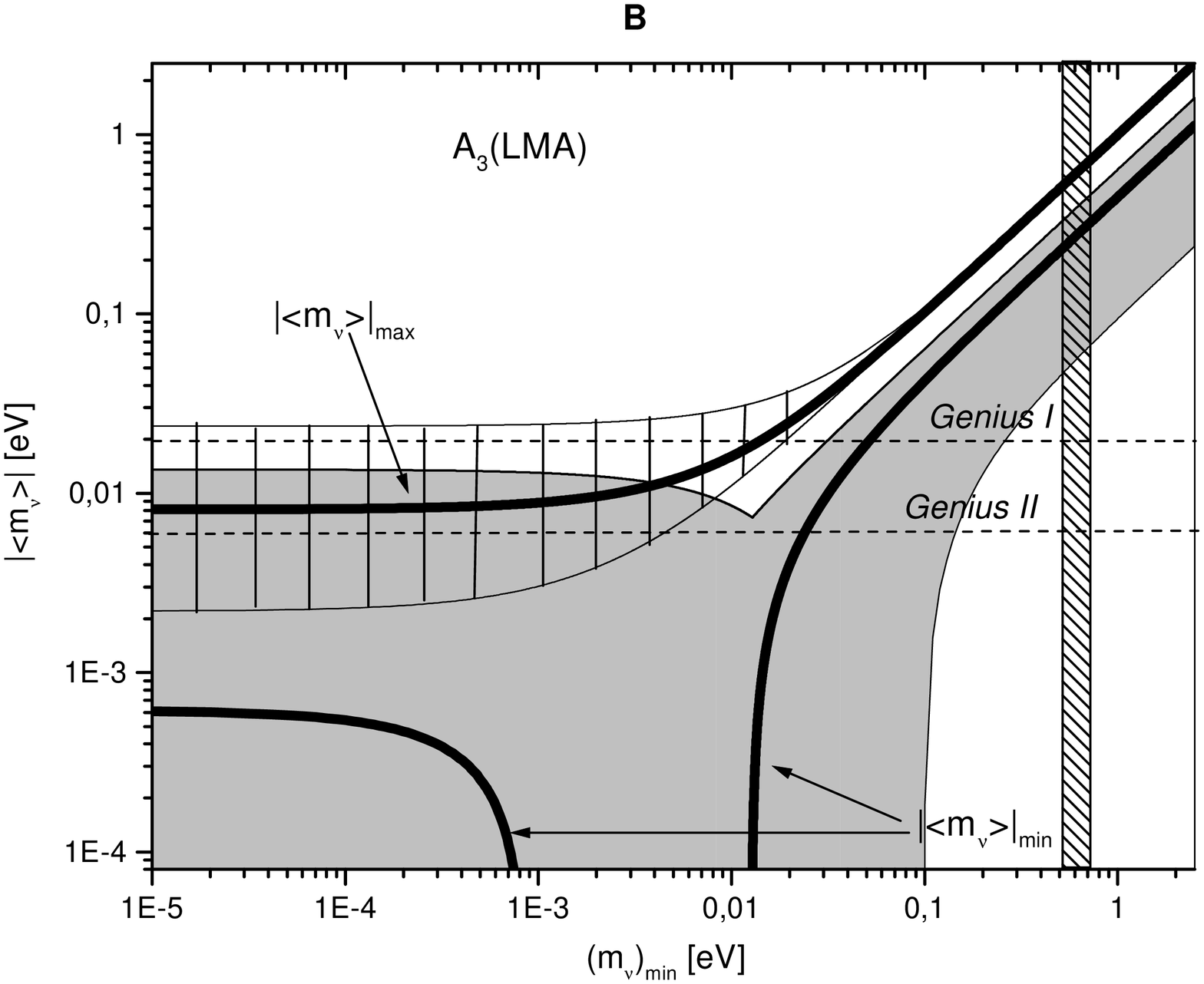,width=8.5cm}
\caption{The values of $|\langle (m_\nu) \rangle |_{max}$ and $|\langle (m_\nu) \rangle |_{min}$ for two solutions
of the solar neutrino problem SMA MSW (left, A) and LMA MSW (right, B) given as function of $(m_\nu)_{min}$ for the best fit
parameters (solid lines). The shaded and hashed regions correspond to the allowed ranges of oscillation parameters (from
Table~\ref{first}) for $|\langle (m_\nu) \rangle |_{min}$ and $|\langle (m_\nu) \rangle |_{max}$ respectively. The
experimatal bounds on $|\langle (m_\nu) \rangle |$ planed by GENIUS I and GENIUS II are depicted. The vertical band corresponds
to the possible value of $m_\beta$ in future $^3_1 H$ decay experiments.
\label{ranglemnumin}}
\end{figure}

\begin{figure}[t]
\epsfig{figure=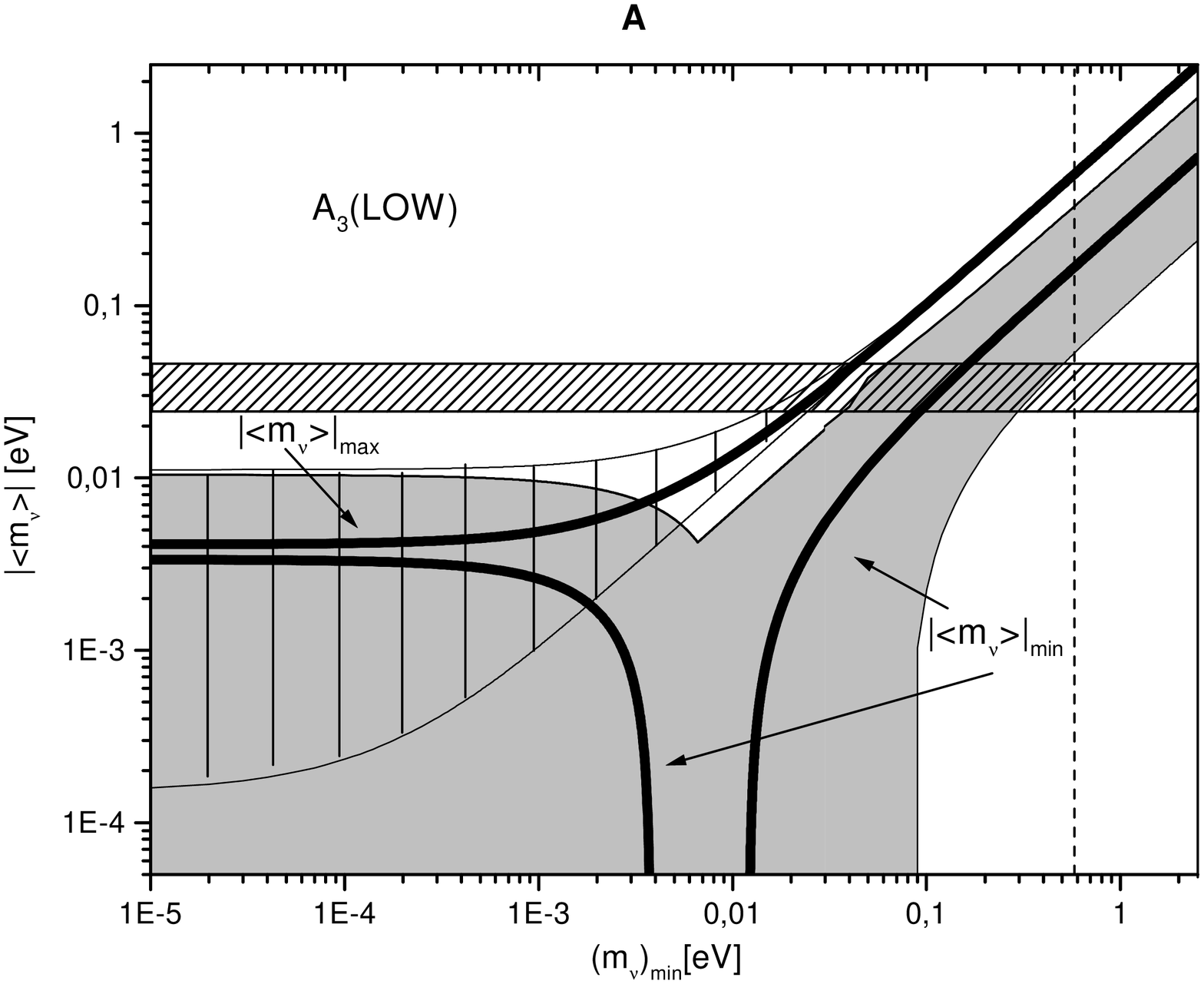,width=8.5cm}
\epsfig{figure=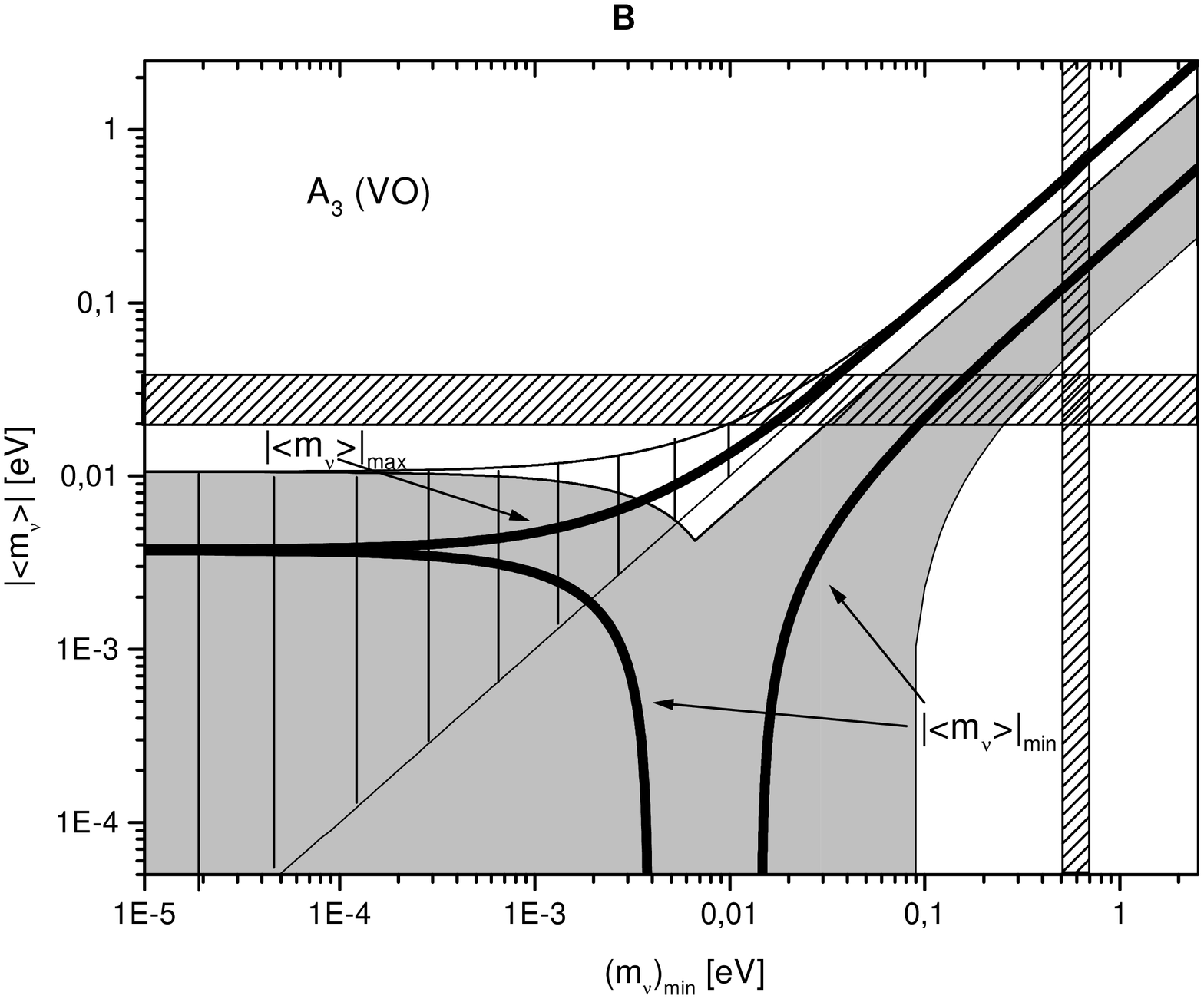,width=8.5cm}
\caption{$|\langle (m_\nu) \rangle |_{max}$ and $|\langle (m_\nu) \rangle |_{min}$ for two solutions of the solar neutrino
problem LOW MSW (left, A) and VO (right, B) as function of $(m_\nu)_{min}$ for the best fit parameters (solid lines).
The shaded and hashed regions correspond to the allowed ranges of oscillation parameters (from
Table~\ref{first}) for $|\langle (m_\nu) \rangle |_{min}$ and $|\langle (m_\nu) \rangle |_{max}$ respectively. In A, the possible
band for $|\langle (m_\nu) \rangle |\in (0.02-0.05)eV$ and the on $(m_\nu)_{min}< 0.6eV$ are depicted. In B, the bands from the two
experiments $(\beta\beta)_{0\nu}$,  $|\langle (m_\nu) \rangle |\in (0.02-0.05)eV$ (horizontal) and $^3_1 H$, $m_\beta \in (0.6-0.8)eV$
(vertical) are given.
\label{boundfig}}
\end{figure}

We see that the bound on the effective mass $|\langle m_\nu \rangle |$
given by the present $(\beta\beta)_{0\nu}$ experiments restricts the
range of possible Majorana neutrino masses. This range depends on the
maximal value of $\sin^2{2\theta}$, and for
\begin{equation}
\sin^2{2\theta} > \frac{1-2|U_{e3}|^2}{(1-|U_{e3}|^2)^2}
\end{equation}
is the same as for Dirac neutrinos \cite{nasze2}.
\\ \\
{\bf Future Perspectives}
\\ \\
Future improving results of neutrino experiments will give better
information on neutrino mass spectra. The improvements that
are important for our purpose and are realistic are the following
\begin{itemize}

\item Concerning neutrino oscillations \cite{exper}
\begin{itemize}
\item the problem of the sterile neutrino should be solved
\item a single solution of the solar neutrino problem should be found
and the value of $\sin^2{2\theta}$ should be given with better
precision
\item the allowed range of $\delta m^2_{atm}$, $\delta m^2_{sol}$ and
$|U_{e3}|$ will be reduced.
\end{itemize}

\item Concerning $(\beta\beta)_{0\nu}$ decay \cite{klapdor}
\begin{itemize}
\item there are plans of going down with the effective Majorana mass
down to $|\langle m_\nu \rangle |\approx 0.006 eV$ in two stages, first
$|\langle m_\nu \rangle |\approx 0.02 eV$ (GENIUS I) and later
$|\langle m_\nu \rangle |\approx 0.006 eV$ (GENIUS II).
\item two possibilities should be envisaged with different impact on
this study. The pessimistic option is that the decay of $^{76}Ge$ will
not be observed and only a new bound on $|\langle m_\nu \rangle |$
will be found. The optimistic, the decay is discovered and a value
$|\langle m_\nu \rangle |\in (0.2 - 0.006) eV$ is inferred.
\end{itemize}

\item Concerning the $^3_1H$ decay \cite{zuber}
\begin{itemize}
\item two Collaborations \cite{weinheimer}, \cite{lobashev}, \cite{zuber} plan
to go down with the value of $m_{\beta}$ below $1 eV$ and perhaps even down
to $0.6 eV$.
\item Once more two scenarios can happen. A distortion in the electron
kinetic energy will not be found and a new bound will follow, $m_{\beta} <
0.6 eV$. On the optimistic side, such a distortion
will be observed, the problem with $m_{\beta}^{2}<0$ will be solved
and the tritium $\beta$ decay will give a value of $m_{\beta}= \kappa' \in
(0.6-2.5) eV$.
\end{itemize}
\end{itemize}
We will now discuss four different possibilities (i) two bounds $|\langle m_{\nu}\rangle | < 0.006 eV$ and $m_{\beta} < 0.6eV$ 
exist, (ii) $|\langle m_{\nu}\rangle | < 0.006 eV$ and $m_{\beta} \approx \kappa'$, (iii) $|\langle m_{\nu}\rangle | \approx \kappa$ 
but $m_{\beta} < 0.6eV$, and finally (iv) $|\langle m_{\nu}\rangle | \approx \kappa$ and $m_{\beta} \approx \kappa'$.
\\ \\
Ad. (i) $|\langle m_{\nu}\rangle | < 0.006 eV$,  $m_{\beta} < 0.6eV$
\\ \\
Nothing special happens, the accepted range of Dirac and Majorana neutrino masses will become smaller. The bound on $m_{\beta}$ gives 
the possible range of Dirac neutrino masses
\begin{equation}
 0.06 eV < (m_{\nu})_{max} \leq 0.6 eV.    \label{eq:16}     
\end{equation}
The bounds on $|\langle m_{\nu}\rangle |$ (depicted in Fig.~\ref{a3} and Fig.~\ref{a3inv}) give a small space of Majorana neutrino 
masses. Once more the 
maximal value of $(m_{\nu})_{min}$ depends on $(sin^{2}2\theta)_{max}$ and $|U_{e3}|^2$, 
which at that time should be known much better. Taking as an 
example for $|\langle m_{\nu}\rangle | < 0.006 eV$ with $|U_{e3}|^2 = 0.01$, only two values of $(sin^{2}2\theta)_{max}$
\begin{eqnarray}
&&(sin^{2}2\theta)_{max} \approx 0.01 (SMA) \Rightarrow (m_{\nu})_{min} < 0.02 eV
\Rightarrow 0.06eV \leq (m_\nu)_{max} \leq 0.062 eV ,\\
&&(sin^{2}2\theta)_{max} \approx 0.98 (LMA) \Rightarrow (m_{\nu})_{min} < 0.05 eV
\Rightarrow 0.06eV \leq (m_\nu)_{max} \leq 0.078 eV .
\end{eqnarray}
We see that a much smaller range of Majorana neutrino masses will be accepted than in the Dirac case. The scheme $A_{3}^{inv}$ will 
be excluded for SMA (LMA) by GENIUS I (GENIUS II) (see Fig.~\ref{a3inv}). 
This is a very pessimistic scenario because the problem of neutrino nature will not 
be solved and the chance to find the spectrum of Majorana neutrinos in a close future will be very small.
\\ \\
Ad. (ii)$|\langle m_{\nu}\rangle | < 0.006 eV$ and $m_{\beta} \approx \kappa'$
\\ \\
If a value of $m_{\beta} \approx \kappa'$ is found the situation changes considerably. With the oscillation parameters and  
$m_{\beta} \approx \kappa'$ we can calculate the spectrum of neutrinos. The only accepted scheme is $A_{3}$. It gives
\begin{eqnarray}
\label{mass1}
m_{1}&=&(m_{\nu})_{min}=[\kappa'^{2}-(1-|U_{e1}|^{2})\delta m^{2}_{solar}-|U_{e3}|^{2}\delta m^{2}_{atm}]^{1/2},
\end{eqnarray}
and 
\begin{eqnarray}
m_{2}&=&[(m_{ \nu})^{2}_{min}+\delta m^{2}_{solar}]^{1/2}, \label{mass2} \\ \label{mass3}
m_{3}&=&[(m_{\nu})^{2}_{min}+\delta m^{2}_{solar}+\delta m^{2}_{atm}]^{1/2}.
\end{eqnarray}
In Fig.~\ref{mnumin} we show the $(m_{\nu})_{min}$ as function of $m_{\beta}
\approx \kappa'$ where the 
oscillation parameters change within their allowed range given in
Tab.~\ref{first}. For $(m_{\nu})_{min} \geq 0.1 eV$,
practically $(m_{\nu})_{min} \approx \kappa'$. The difference is
visible only for very small $(m_{\nu})_{min}$. This means
that the experimental error on $\kappa'$ is the only significant source of the variation
range of $(m_{\nu})_{min}$. The mass 
spectrum obtained from Eq.~\ref{mass1},~\ref{mass2} and~\ref{mass3} does not depend on the neutrino
nature. If neutrinos are Majorana 
particles, the value of $(m_{\nu})_{min}$ obtained from $H^{3}_{1}$ decay
can be used to find the range of 
possible values of $|\langle m_{\nu}\rangle |$. In Fig.~\ref{ranglemnumin} we depicted 
$|\langle m_{\nu}\rangle |_{max}$ and  $|\langle m_{\nu}\rangle |_{min}$ for two 
possible solutions of the solar neutrino problem,  SMA (Fig.~\ref{ranglemnumin}A) and LMA (Fig.~\ref{ranglemnumin}B) 
in double logarithmic scales.
The bound of  $(m_{\nu})_{min} \sim (0.6-0.8) eV$ gives the
range of possible $|\langle m_{\nu}\rangle |$ values 
which follows from crossing the space allowed by oscillation experiments.
\begin{equation}
|\langle m_{\nu}\rangle |_{\beta}^{min} \leq |\langle m_{\nu}\rangle | \leq |\langle m_{\nu}\rangle |_{\beta}^{max}, \label{eq:19}
\end{equation}
If the experimental bound on $|\langle m_{\nu}\rangle |$ from $(\beta
\beta)_{0\nu}$ decay is below the $|\langle m_{\nu}\rangle |_{\beta}^{min}$ range, then
neutrinos can not be Majorana particles. If it is lager, the problem of
neutrino nature is not solved.
\\ \\
Ad.(iii) $|\langle m_{\nu}\rangle | \approx \kappa$, $m_{\beta} < 0.6 eV$.
\\ \\
It is quite probable that this scenario will happen. In this case
the neutrinos are Majorana 
particles. In Fig.~\ref{boundfig}A a possible band of $|\langle m_{\nu}\rangle |$, obtained
from the GENIUS experiment is given. This 
bound crosses the region of space allowed by oscillation data giving the possible value of $(m_{\nu})_{min}$
\begin{equation}
(m_{\nu})_{min}^{min(\beta \beta)_{0\nu}}\leq (m_{\nu})_{min}\leq
(m_{\nu})_{min}^{max(\beta \beta)_{0\nu}}, 
\label{eq:20}
\end{equation}
If the value of $|\langle m_{\nu}\rangle |$ is found in the second stage of the GENIUS
experiment $|\langle m_{\nu}\rangle | \approx 0.01-0.005 eV$
the $(m_{\nu})_{min}^{min(\beta \beta)_{0\nu}} = 0$. With the range of
possible $ (m_{\nu})_{min}$ the 
result on $m_{\beta}$ from tritium $\beta$ decay can
predicted. In practice $m_{\beta}$ should also satisfy 
the inequalities given by Eq.( \ref{eq:20}). If the experimental
limit on $m_{\beta}$ is larger than 
$(m_{\nu})_{min}^{max(\beta \beta)_{0\nu}}$, the theory with three Majorana neutrinos is consistent. If,
what is less probable, the experimental limit on $m_\beta$ is smaller than 
$(m_{\nu})_{min}^{max(\beta \beta)_{0\nu}}$ then the theory with three Majorana neutrinos is ruled out.
\\ \\
Ad.(iv)  $|\langle m_{\nu}\rangle | \approx \kappa$, $m_{\beta} \approx \kappa'$.
\\ \\
This is the best, but on the other side, the least probable
scenario. The value of 
$|\langle m_{\nu}\rangle | \neq 0$ defines the neutrino as a Majorana particle. The
value $m_{\beta} \neq 0$ gives their mass 
spectrum. Comparing both bands with the region of $|\langle m_{\nu}\rangle |$
allowed by the oscillation data (Fig.~\ref{boundfig}B) is a 
check of internal consistency of the theory. With precise data
the crossing of the three regions can be used
to specify the values of the CP breaking Majorana phases $
\varphi_{1}$ and $ \varphi_{2}$ (Eq.(\ref{basic})). 
If the two bands $ |\langle m_{\nu}\rangle |$ and $m_{\beta}$ cross the oscillation
region near $|\langle m_{\nu}\rangle |_{max}$, two 
phases are equal $ \varphi_{1}=\varphi_{2}\approx n \pi$. This means
that all three Majorana neutrinos have the 
same CP parity $\eta_{CP}=+i$ and the symmetry is conserved. If the two
bands cross the oscillation region 
near $|\langle m_{\nu}\rangle |_{min}$ once more the CP symmetry is satisfied with  
$\eta_{CP}(\nu_{1})=-\eta_{CP}(\nu_{2})=-\eta_{CP}(\nu_{3})=i$.
Finally, if all three regions cross somewhere 
in between, the phases $ \varphi_{1}$ are nontrivial and the CP symmetry is broken.
\\ \\
{\bf In conclusion}
\\ \\
Present experimental data define the region of neutrino mass. The
heaviest Dirac neutrino mass must be 
somewhere in the range $0.06 eV \leq (m_{\nu})_{max} \leq 2.5 eV$.
The analogous range for Majorana neutrinos depends on the solution of the solar neutrino problem
and is $0.06 eV \leq (m_{\nu})_{max} \leq 0.26 eV$ for SMA and $0.06 eV \leq (m_{\nu})_{max} \leq 1.5 eV$
for the LMA scenario. A better future bound on the 
effective mass in the neutrinoless double $\beta$ decay and the
tritium $\beta$ decay would give 
better limits on Dirac and Majorana neutrino masses but the problem of the
mass spectrum and their nature
will still not be solved. The situation will change significantly if at least one of the experiments
gives a positive result and $|\langle m_\nu \rangle |$ or $m_\beta$ will be different from zero. The most
difficult scenario, where both experiments give positive results is the one where (i) the nature of neutrinos,
(ii) their mass spectrum, (iii) the consistency of the theory and (iv) the CP breaking Majorana phases can
be found.
\\ \\
{\bf Acknowledgments}
This work was supported by the Polish Committee for Scientific Research under Grant No. 2P03B05418 and 
2P03B04919, by CICYT and by Junta de Andalucia.

\end{document}